\documentclass[prb,twocolumn,floatfix,amsfonts,showpacs,superscriptaddress]{revtex4-1}
\usepackage{graphicx,graphics,color,epsfig}
\usepackage{bm}
\usepackage{amsmath}
\usepackage{amssymb}

\begin{document}

\title{First principles investigation of pressure related quantum transport in pure black phosphorus devices}

\author{Ximing Rong}
\affiliation{Shenzhen Key Laboratory of Advanced Thin Films and Applications, College of Physics and Energy, College of Electronic Science and Technology, Shenzhen University, Shenzhen, 518060, China}
\affiliation{Department of Physics and Shenzhen Institute of Research and Innovation, the University of Hong Kong, Pokfulam Road, Hong Kong SAR, China}

\author{Zewen Wu}
\affiliation{School of Physics, Beijing Institute of Technology, Beijing 100081, China}
\affiliation{Department of Physics and Shenzhen Institute of Research and Innovation, the University of Hong Kong, Pokfulam Road, Hong Kong SAR, China}

\author{Zhizhou Yu}
\affiliation{School of Physics and Technology, Nanjing Normal University, Nanjing 210023, China}

\author{Junjun Li}
\affiliation{Hongzhiwei Technology (Shanghai) Co., Ltd., Shanghai 201206, China}

\author{Xiuwen Zhang}
\affiliation{Shenzhen Key Laboratory of Advanced Thin Films and Applications, College of Physics and Energy, College of Electronic Science and Technology, Shenzhen University, Shenzhen, 518060, China}

\author{Bin Wang}
\email{binwang@szu.edu.cn}
\affiliation{Shenzhen Key Laboratory of Advanced Thin Films and Applications, College of Physics and Energy, College of Electronic Science and Technology, Shenzhen University, Shenzhen, 518060, China}

\author{Yin Wang}
\email{yinwang@shu.edu.cn}
\affiliation{Department of Physics and Shenzhen Institute of Research and Innovation, the University of Hong Kong, Pokfulam Road, Hong Kong SAR, China}
\affiliation{Department of Physics and International Centre for Quantum and Molecular Structures, Shanghai University, 99 Shangda Road, Shanghai 200444, China}

\begin{abstract}
We propose a first-principles calculation to investigate the pressure-related transport properties of two kinds of pure monolayer black phosphorus (MBP) devices. Numerical results show that semi-conducting MBP can withstand a considerable compression pressure until it is transformed to be a conductor. The pure MBP devices can work as flexible electronic devices, "negative" pressure sensors, and "positive" pressure sensors depending on the chirality of BP and the magnitude of vertical pressure. When pressure is relatively small, the conductance is robust against the stress for zigzag MBP devices, while shows pressure-sensitive properties for armchair MBP devices. The pressure-stable property of zigzag MBP devices implies a good application prospects as flexible electronic devices, however, the distinct negative increase of conductance versus pressure indicates that armchair MBP devices can work as "negative" pressure sensors. When pressure is relatively large, both armchair MBP devices and zigzag MBP devices show favorable properties of "positive" pressure sensors, whose conductivities rise promptly versus pressure. The longer the device, the more the pressure sensitivity. Band alignment analysis and empirical Wentzel$-$Kramers$-$Brillouin (WKB) approximations are also performed to testify the tunneling process of pure MBP devices from first principles calculation.
\end{abstract}

\maketitle
\section{Introduction}

Two-dimensional (2D) materials, including graphene \cite{Novoselov,Morozov,lee1,Novoselov2,Qiao,Bin,bin2,JW}, transition metal dichalcogenide \cite{mak3,gomez1,mak2, mak1,zl,zyf}, silicene \cite{vogt1,houssa1,arfune}, and black phosphorus (BP) \cite{li1, churchill1,rodin1,peng1,liang1,wang3, tran1, liu1, qiao1,mingyan,bowen,mingyan2,xiao,Pablo,koda,deniz}, have attracted intense interests in material science and quantum physics due to their remarkable physical and chemical properties. Among these materials, BP shows significant application prospect in electronic devices due to the relatively high carrier mobility and adjustable band gap \cite{li1,churchill1}. Due to the puckered configuration, BP is much easier to realize structural deformation in all three dimensions by tension or compression comparing to planar structures, such as graphene and silicene, and to accomplish large-scale band gap modulation. Rodin \emph{et al.} \cite{rodin1} predicted the band structures of strain-related monolayer BP (MBP) using tight-binding model and density functional theory (DFT), and indicated that a deformation normal to the 2D plane can change the band gap and induce a semiconductor-metal phase transition of MBP. Peng \emph{et al.} \cite{peng1} investigated the band structure of MBP under in-plane strain using DFT method accompanied with hybrid functional, and pointed that MBP can withstand a tensile stress as high as 10 N/m and a strain up to 30$\%$. They found that axial strain can induce a phase transition with direct-indirect-direct band gap. The favorable strain sensitivity and strain endurance make BP an ideal material for strain-sensing electronics and flexible electronic devices.

So far, many researches have been carried out to investigate the pressure-induced quantum response in BP based nanoscale devices experimentally and theoretically \cite{xiao,Pablo,koda,deniz}.  Xiao \emph{et al.} manipulated few-layer BP nanosheet by chemical vapor transport method \cite{xiao}, and observed a phase transition from orthorhombic semiconductors to simple cubic metal with increasing pressure by performing in situ ADXRD and Raman spectroscopy with the assistance of DAC apparatus. They also carried out first principles calculation to interpret the metallic behaviors of BP under pressure. Pablo $\emph{et al.}$ investigated the funnel effect in MBP \cite{Pablo}, which describes the possibility of controlling exciton motion by means of inhomogeneous strains. They found that funnel effect in BP is much stronger than that in $MoS_2$, and more crucially, shows opposite behaviors as that in $MoS_2$. Excitons in BP are mainly accumulated isotropically in strain-reduced incompact regions, instead of occurring in the regions with high tensile strain like in MoS2. Deniz \emph{et al.} investigated the strain-related optical properties of MBP using first principles calculation \cite{deniz}, and found that the optical response of MBP are sensitive to the magnitude and the orientation of the applied strain due to the strong anisotropic atomic structure of BP.    Based on first-principles calculations, Koda \emph{et al.} studied the electric behaviors of BP-$MoSe_2$ and BP-$WSe_2$ hetero-bilayers, and analyzed the long-range structural bending affecting to electronic properties due to Wan der Waals interaction \cite{koda}. Despite lots of interests in pressure response of BP based nanodevices, the strain-related quantum transport in pure MBP devices has so far been rarely explored. This is the purpose of this manuscript to try to provide an idea to fertilize this field.

In this manuscript, we build two kinds of pure MBP nano-devices and investigate their pressure-related electric properties using the first principles calculation method. Comparing to other metal-MBP hetero-junctions, pure MBP device has more advantages including simple structure, easy preparation, smooth continuous interface, and lattice mismatch avoidable, and so on. We want to address the following questions of pure MBP devices. 1) How does the chirality of pure MBP devices influence the pressure-related quantum transport behaviors? Is either the zigzag or the armchair device proper to be a pressure sensor devices? 2) How does the magnitude of pressure influence the transport properties of pure MPB devices? 3) What is the length dependence of conductance of pure MBP devices? 4) How does structure relaxation influence the physical performance of MBP pressure sensor? To answer these questions, first principles calculation were carried out to investigate the quantum transport of pure MBP devices within the framework of combination of non-equilibrium Green's function (NEGF) and density functional theory (DFT) \cite{taylor1,taylor2}.

The subsequent part of this manuscript is organized as follows. In the second part, we describe our pure MBP nano-devices and introduce the first principles modelling methods. In the third part, we show the numerical results and physical analysis about pressure-related MBP devices in several aspects including mechanical, electric, transport, and band alignment. The numerical result about tunneling probability is also fitted by classical WKB approximation to indicate the reliability of our first principles calculation. Finally, we give a conclusion of this manuscript.

\begin{figure}[t]
\includegraphics[width=8cm]{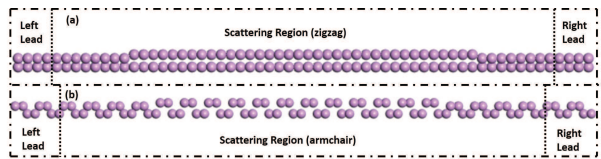}
\caption{(Color online) Schematic structures of (a) zigzag MBP device with length of central compression-tunable MBP section equal to 18$L$, and (b) armchair MBP device with length of central compression-tunable MBP section equal to 12$L$.
}
\label{fig1}
\end{figure}

\section{Simulation details}
Fig.1(a) and (b) show two different pure MBP nano-devices, in which quantum transport is along zigzag direction and armchair direction of MBP, respectively. For each device,
its two probes are formed by compressed MBPs with fixed compression ratio equal to 30$\%$ to ensure the conducting behaviors as will be discussed in the following sections, and
the central region is composed of a section of MBP with tunable compression ratio from zero to 30$\%$. In this manuscript, compression ratio is defined as $R_C = (1-h/h_0)\times 100\%$ , where $h_0$ and $h$ represent the thickness of the free standing and compressed MBP along vertical direction to the MBP plane, respectively. The length of central region is equal to 2$L$, 4$L$, 6$L$, 12$L$, 18$L$, and 24$L$, where $L$ represents the length of a periodic unit cell of MBP along transport direction. Both zigzag and armchair MBP devices as shown in Fig.1(a) and (b) are periodic in perpendicular direction to quantum transport in the MBP plane. We want to emphasize that the size of the largest structures with central region equal to 24$L$ is already compatible with the practical scale of transistor, where the length $L \approx$ 7.9nm for zigzag MBP device and $L \approx$ 11.1nm for armchair MBP device.

In our investigation, two different DFT codes, VASP \cite{kresse1,kresse2} and Nanodcal \cite{taylor1,taylor2}, were used to model the electric behavior and quantum transport of pressure-related MBP devices, respectively. The structural relaxations of all the compressing procedures of MBP were implemented by VASP, and accomplished by Perdew-Burke-Ernzerhof (PBE) functional along with PAW potentials \cite{blochl1,perdew1}. The kinetic energy cutoff was chosen to be 500 eV and the reciprocal space was meshed by $13 \times 9 \times 1$ using the Monkhorst-Pack method \cite{monkhorst1}. The volumes of structures were fully relaxed until the atomic force is smaller than 0.001 $eV/{\AA}$. The relaxed lattice constants for free standing MBP unit cell are 4.625 ${\AA}$ in armchair direction, 3.298 ${\AA}$ in zigzag direction, and 2.102 ${\AA}$ in vertical direction between two nonequivalent P atomic layers, which are in good agreement with the recognized DFT results \cite{peng1}.

The quantum transport properties of MBP devices were implemented by the transport package Nanodcal, which is based on the standard NEGF-DFT method \cite{taylor1,taylor2}. In our calculation, norm-conserving non-local pseudo-potential was used to define the atomic cores \cite{troullier1}, and atomic orbital basis set with single-$\zeta$ plus polarization was used to expand physical quantities \cite{soler1}. The exchange correlation potential was treated using the PBE functional \cite{blochl1,perdew1}. Finally, the NEGF-DFT self-consistency was carried out until the numerical tolerance of the Hamiltonian matrix is less than $10^{-4}$ eV.

\begin{figure}[t]
\centering
\includegraphics[width=8cm]{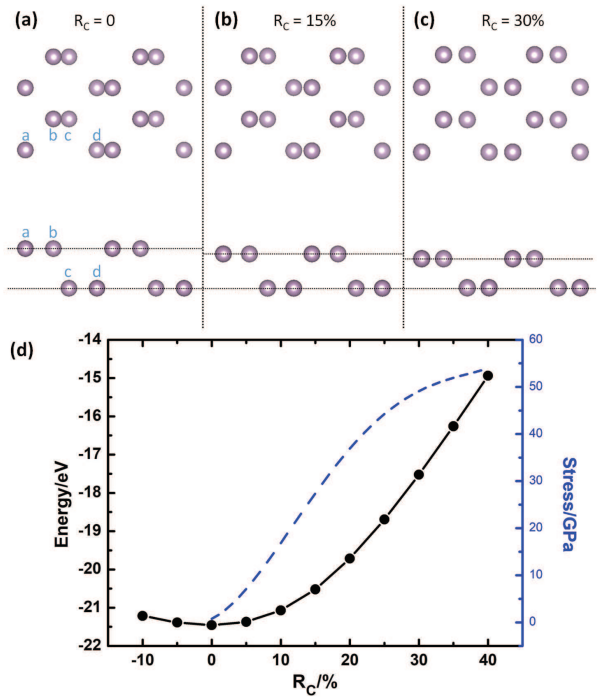}
\caption{(Color online) (a)-(c) Top views and side views of the partially relaxed MBP structures with different compressing ratio $R_C$. (d) Total energy (black solid circle curve) and stress (blue dash curve) of a MBP unit cell as a function of compression ratio $R_C$.
}
\label{fig2}
\end{figure}

\section{Results and discussion}

In this section, we show our numerical investigation of mechanical and electric properties of pressure-related periodic MBP, and then quantum transport and band alignment analysis of pure MBP nanoscale devices.

\subsection{Mechanical properties of pressure-related MBP}
In this section, we discuss the mechanical behaviors of periodic 2D MBP with the increase of pressure along vertical direction to the 2D plane. Normally, a vertical compression could cause the expansion of MBP in both zigzag and armchair directions. Nevertheless, in most of ours investigation, we suppose the pressure induced deformation mainly occurs along the armchair direction of the MBP nano-devices, but only a few change along the zigzag direction. So in our first principles calculation, the size of box can be relaxed along armchair direction under different pressure, but fixed along zigzag direction. This assumption is necessary in quantum transport investigation for two dimensional devices using DFT method, where the size of periodic unit cell along transverse direction should be fixed. This preference is also reasonable considering of the anisotropic mechanical properties of MBP in zigzag and armchair directions. Previous investigation indicated that P atoms are more preferred to move along the armchair direction but not zigzag direction under vertical pressure because the Young's modulus of MBP in armchair direction (44 GPa) is much smaller than that in zigzag direction (166 GPa) \cite{rodin1}. In the following part of this manuscript, this kind of structures is called as partially relaxed structures, and most of our investigation is based on this kind of relaxation. In addition, we also built some fully relaxed structures as comparison to check the accuracy of our numerical results, where "fully" means both the zigzag direction and armchair direction of MBP can be relaxed inside a size-tunable box under different compression ratio.

Fig.2(a)-(c) present the top views and side views of partially relaxed MBP with $R_C$ equal to zero, 15$\%$, and 30$\%$, respectively, where $\emph{a}$, $\emph{b}$, $\emph{c}$, and $\emph{d}$ represent P atoms in a unit cell. With the increase of compression ratio from zero to 30$\%$, the distance between atom $\emph{a}$ and atom $\emph{b}$ (or atom $\emph{c}$ and atom $\emph{d}$) in the same layer got decreased from 1.49 ${\AA}$ ($R_C$ = 0) to 1.11 ${\AA}$ ($R_C$ = 0.3), while the distance between atom $\emph{b}$ and atom $\emph{c}$ increases from 0.83 ${\AA}$ to 1.2 ${\AA}$. Meanwhile, the distance between atom $\emph{a}$ and atom $\emph{b}$ (or atom $\emph{c}$ and atom $\emph{d}$) alongside zigzag direction remains unchanged as 1.65 ${\AA}$.

To estimate the mechanical stress during this compressing procedure, we calculated the stress-pressure relationship of MBP as a function of compression ratio $R_C$ using the method described in the references \cite{roundy1,luo1,rong1}, where the stress can be calculated by $-\partial E/\partial d*1/S$ with $d$ the thickness of MBP and $S$ the pressed area. As shown in Fig.2(d), pressure energy is presented by the black solid-dot curve, and stress is presented by the blue dash curve. With the increase of $R_C$ from zero to 40$\%$, pressure energy increases smoothly from -21.4 eV to -14.8 eV. In addition, pressure energy also increases when $R_C$ decreases from zero to -10$\%$, indicating the stability of MBP under zero pressure. The smooth increase of pressure energy means that MBP can always maintain its arrayed structure regularly under large compression, and do not appear any structural defect or bonding twist. Stress-stain curve shows that MBP sustains higher stress by larger compression ratio, which is reasonable because atoms will be harder to get closer due to the intermolecular repulsion and is inevitable to increase the pressure load to accomplish the acquired compressing pressure. By linear fitting the stress curve versus compression ratio from $R_C$ = 0 to $R_C$ = 5$\%$, the Young's modulus vertical to MBP plane can be obtained equal to 127 GPa.

\begin{figure}[t]
\includegraphics[width=8cm]{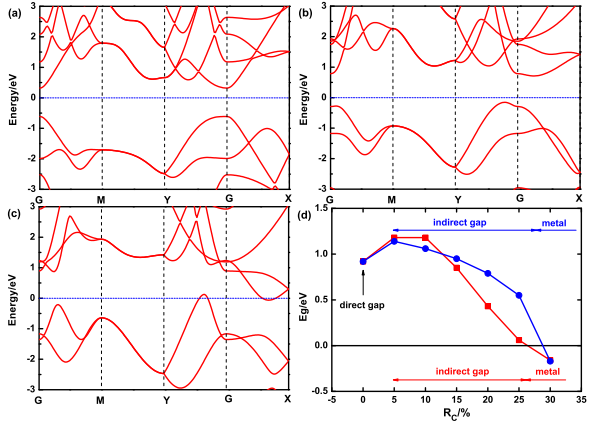}
\caption{(Color online) (a)-(c) Band structures of partially relaxed MBP with compression ratio equal to zero, 15$\%$ and 30$\%$, respectively. (d) Band gap as a function of $R_C$ for partially relaxed (blue circle broken-line) and fully relaxed (red quadrate broken-line) MBP structure. The black horizontal line at $E_g$=0 indicates the Fermi level.
}
\label{fig3}
\end{figure}

\subsection{Electric properties of pressure-induced MBP}
In this section, we show the electric behaviors of pressure-induced MBP. Fig.3(a)-(c) plot the band structures of partially relaxed 2D MBP with compression ratio equal to zero, 15$\%$, and 30$\%$, respectively. For the free standing MBP, a direct band gap appears at $\Gamma$ point and is roughly equal to 0.91 eV, which is in agreement with several theoretical works elsewhere \cite{liu1,peng1,tran1}. Despite this gap is smaller than that from experiments \cite{liang1,wang3} and GW method \cite{tran1}, the substance of pressure related quantum transport in pure MBP devices in this paper is not influenced. When $R_C$ increases from zero to 15$\%$, MBP transforms from a direct band gap material to an indirect band gap material as shown in Fig.3(b). While, when $R_C$ is further increased to 30$\%$ as shown in Fig.3 (c), the conduction band minimum (CBM) has descended below the valance band maximum (VBM), and MBP is finally changed to be a conductor.

Fig.3(d) shows the variation of band gap of partially relaxed MBPs as a function of compression ratio, which is shown by blue solid-circle curve. As a comparison, band gap variation of fully relaxed MBPs is also plotted by red solid-square curve. We try to explore the influence of structure relaxation to the physical essentials of electric behavior. For both structures, the band gap increases firstly versus $R_C$, and then changes from a direct one to an indirect one with $R_C$ roughly equal to 5$\%$. With further increase of $R_C$, the band gap decreases continuously and MBP eventually becomes a conductor when $R_C$ is roughly equal to 25$\%$-30$\%$. The fully relaxed MBP and partially relaxed MBP show qualitatively consistent behaviors of gap variation and even the same phase transition points, although the fully relaxed MBP drops dramatically when the pressure increases from $R_C$ = 10$\%$ to $R_C$ = 25$\%$.

\begin{figure}[t]
\includegraphics[width=8cm]{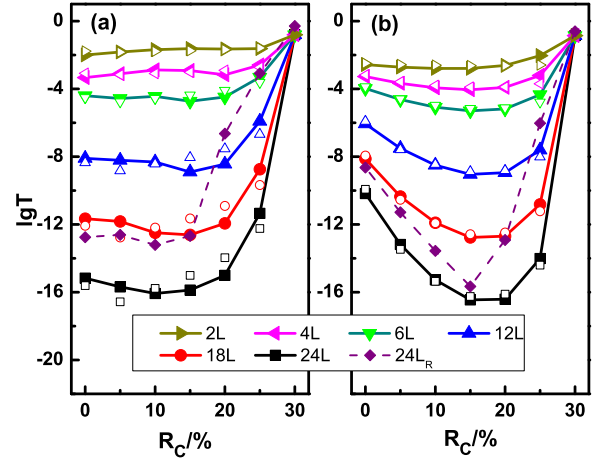}
\caption{(Color online) $T(E)$ (colored solid symbol curve) and $T_{WKB}$ (hollow scatterplots) at the Fermi level in logarithmic scale versus $R_C$  for (a) partially relaxed zigzag MBP devices, and (b) partially relaxed armchair MBP devices with different length. The dash curves with diamond dots in both panels represent $T(E_F)$ of fully relaxed zigzag and armchair MBP devices with length equal to 24$L$.
}
\label{fig4}
\end{figure}

\subsection{Pressure-related quantum transport of MBP structures}
In this subsection, we focus on the relationship between transmission coefficient and pressure of pure MBP devices. The transmission coefficient of two probe system can be calculated by the following equation within the NEGF method \cite{bgwang},
\begin{equation}
T(E)=Tr(\Gamma_L G^R \Gamma_R G^A),
\end{equation}
where $G^R$ and $G^A$ are the retarded and advanced NEGFs of the system, respectively, and $\Gamma_L$ and $\Gamma_R$ are the line-width functions describing the interaction between leads and scattering. Fig.4(a) and (b) show the length- and compression- dependent transmission coefficient at equilibrium state for the zigzag pure MBP devices (see Fig.1(a)) and armchair pure MBP devices (see Fig.1(b)), respectively. For each device, $R_C$ is fixed equal to 30$\%$ for both MBP leads, while changes from zero to 30$\%$ for the central MBP section in the scattering region.

For the zigzag MBP devices, several information can be found as shown in Fig.4(a). Firstly, the longer the structure, the smaller the transmission coefficient. This is reasonable because a longer structure corresponds to a higher potential barrier in the scattering region due to the semiconducting behavior of MBP section in the central region. Secondly, $T(E)$ is nearly invariable for all the structures with different length when $R_C$ is smaller than 20$\%$, indicating zigzag MBP devices are pressure-stable under relatively small pressure and have a good application prospects of flexible electronic devices \cite{mingyan,mingyan2,zhizhou}. Thirdly, $T(E)$ increases dramatically after $R_C >$ 20$\%$ for all the structures, and finally all the zigzag MBP devices show good conductivity when $R_C$ is roughly equal to 30$\%$. This means zigzag MBP devices can work as perfect pressure sensors when stress is large enough.

For the armchair MBP devices as shown in Fig.4(b), most of the behaviors of $T(E)$ are similar to those of zigzag devices, except an obvious difference. $T(E)$ experiences a monotonously decrease when $R_C$ increases from 0 to 15$\%$, showing "negative" pressure response, which is very different from the condition of zigzag MBP devices. The longer the structure, the more obvious the decreasing tendency. With further increasing of $R_C$, $T(E)$ increases dramatically and show perfect "positive" pressure response. By comparing Fig.4(a) and (b), we can give such a conclusion. When pressure is small, zigzag MBP devices could be well used as pressure-stable electronic devices, while armchair MBP devices can work as "negative" pressure sensor. When pressure is large, both zigzag and armchair MBP devices are very good "positive" pressure sensors.

\subsection{Band alignment analysis and WKB fitting}
We also performed an empirical WKB calculation to verify our first principles results of transmission coefficients for all the MBP devices. In WKB calculation, the energies of valence band alignment $\Delta E_V$ and conduction band alignment $\Delta E_C$ in the scattering region of two probe MBP structures are needed \cite{wang4,wang5,hou1,politzer1,mao1}, which are defined as
\begin{eqnarray}
\Delta E_V = E_F - E_{VBM}, \nonumber \\
\Delta E_C = E_{CBM} - E_F,
\end{eqnarray}
where $E_F$ is the Fermi energy of the conducting compressed MBP lead; $E_{VBM}$ and $E_{CBM}$ are the valence band minimums (VBM) and the conduction band maximums (CBM) of central thickness-variable MBP section, respectively. To obtain $\Delta E_V$ and $\Delta E_C$, band offsets analysis is performed for all the MBP devices under different $R_C$ by projecting
the density of states (PDOS) in scattering region along the transport direction.

\begin{figure}[t]
\includegraphics[width=8cm]{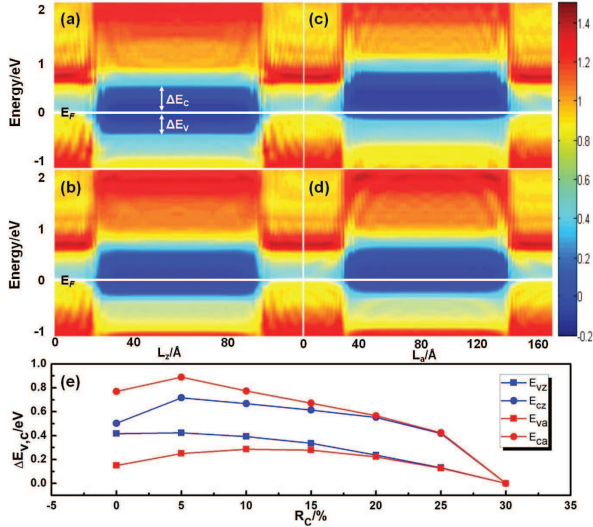}
\caption{(Color online) (a)-(d) Real space distribution of PDOS in logarithmic scale along the transport direction for (a) zigzag MBP device with $R_C=0$, (b) zigzag MBP device with $R_C=15\%$, (c) armchair MBP device with $R_C=0$, and (d) armchair MBP device with $R_C=15\%$. (e) $\Delta E_V$ and $\Delta E_C$ as functions of $R_C$ for zigzag ($E_{cz}$ and $E_{vz}$) and armchair ($E_{ca}$ and $E_{va}$) MBP devices. The length of all the structures are equal to 24$L$.
}
\label{fig5}
\end{figure}

Fig.5(a)-(d) show PDOS of two zigzag MBP devices and two armchair MBP devices with length equal to 24$L$ and compression ratio equal to zero and 15$\%$. For each structure, the Fermi level always in the range of VBM and CBM, and locates closer to VBM, hinting a p-type hole-like transport. 15$\%$ compression ratio can induce an increase of $\Delta E_V$ and a decrease of $\Delta E_C$ for both zigzag and armchair devices. In addition, the band bending can also be observed obviously at the interface of metallic MBP leads and center semi-conducting MBP section due to the difference of their work functions. Another method to analyze the band alignment of tunneling junction is supercell method, which is introduced in detail in Ref. \cite{wang5,yin1}. Fig.5(e) shows a global graph of band offsets $\Delta E_V$ and $\Delta E_C$ versus $R_C$ for zigzag devices and armchair devices with 24$L$ length. When $R_C$ is small, $\Delta E_V$ and $\Delta E_C$ increase monotonous. With increasing of $R_C$, $\Delta E_V$ and $\Delta E_C$ show roughly decreasing behavior for both zigzag structure and armchair structure. When $R_C$ is equal to 30$\%$, $\Delta E_V$ and $\Delta E_C$ are zero because a semiconductor-metal phase transition occurs for the central MBP section. Band alignment analysis reveals accordant information of band gap variation as shown in Fig.3(d), and gives insight view of the quantum transport in pure MBP devices under different pressure.

Based on the obtained $\Delta E_V$ and $\Delta E_C$ under different $R_C$, the transmission coefficient can be estimated empirically by WKB method \cite{wang4,wang5,hou1,politzer1,mao1}, where $T_{WKB}$ can be simply described as
\begin{equation}
ln T_{WKB}(E) \propto -l\times \sqrt{m_{eff}E}.
\end{equation}
Here $l$ is the tunneling distance and $m_{eff}$  is the effective mass of the system. For hole-like tunneling process,  $m_{eff}E$  can be calculated by \cite{cai1,freeman1}
\begin{equation}
\frac{1}{m_{eff}E} = \frac{1}{m_{C}\Delta E_C} + \frac{1}{m_{V}\Delta E_V},
\end{equation}
where $m_V$ and $m_C$ are the effective masses of valence and conduction band of MBP, respectively.

As a comparison, $T_{WKB}$ is also plotted in Fig.4(a) and (b) for all the MBP devices by hollow scatterplots. We can find that the results from WKB method are quantitatively accordant with those from first principles calculation, indicating our numerical results from first principles method in this work is credible.

To further examine the influence of structure relaxation to the physical performance of the pure MBP devices, transmission coefficients of fully relaxed zigzag and armchair MBP devices with length equal to 24$L$ were also calculated and plotted by the diamond-dash curves in Fig.4(a) and (b). For the zigzag devices, $T(E)$ is not sensitive to the pressure when $R_C$ is smaller than 15$\%$. While, for the armchair devices, $T(E)$ shows "negative" pressure response versus $R_C$ when $R_C$ is smaller than 15$\%$. This important result is qualitatively agreement with that obtained from partially relaxed MBP devices. When $R_C$ is larger than 15$\%$, $T(E)$ increases with $R_C$ with a faster speed than the partially relaxed structure for both zigzag and armchair devices. This is reasonable because the band gaps of the fully relaxed structures between $R_C$ = 20$\%$ and 25$\%$ are smaller than those of the partially relaxed structures as shown in Fig.3(d). The consistency of the fully relaxed and partially relaxed results of transmission coefficients show that structural relaxation in MBP devices maybe not so sensitive to influence their pressure-related transport tendency.

\section{Conclusions}
We investigated the pressure-related quantum transport properties of two kinds of pure MBP devices. We found that MBP can sustain relatively large pressure, and occur phase transition from a direct band gap semi-conductor to an indirect band gap semi-conductor, and then a conductor with increasing of compression ratio. The pure MBP devices show good application prospect as pressure sensor. The longer the device, the more sensitive the pressure sensor. When pressure is small, zigzag MBP devices show pressure-stable properties and can work as flexible electronic devices, while armchair MBP devices show pressure-sensitive properties and can work as "negative" pressure sensors with conductance decreasing versus compression ratio. When pressure is large, both armchair MBP devices and zigzag MBP devices can work as "positive" pressure sensors, whose conductivity rise promptly versus pressure. Although the main conclusion is obtained based on the partially relaxed structures, we confirmed that structure relaxation can only quantitatively but not qualitatively influences the quantum transport of pressure-related pure MBP devices.

\section*{Acknowledgments}

This work was financially supported by grants from the National Natural Science Foundation of China (Grant No. 11774238, 11774239 and 11404273), Shenzhen Key Lab Fund (ZDSYS20170228105421966), the University Grant Council (Contract No. AoE/P-04/08) of the Government of HKSAR.

\end{document}